\documentclass[aps,superscriptaddress,twocolumn,a4paper,showpacs]{revtex4}
\usepackage[colorlinks=true, pdfstartview=FitV, linkcolor=blue, citecolor=red, urlcolor=black]{hyperref}
\usepackage{graphicx}
\usepackage[all]{xy}
\usepackage{amsmath}
\usepackage{amssymb}
\usepackage{tensor}
\newcommand{\be}{\begin{equation}}
\newcommand{\ee}{\end{equation}}
\newcommand{\ben}{\begin{eqnarray}}
\newcommand{\een}{\end{eqnarray}}
\newcommand{\bes}{\begin{subequations}}
\newcommand{\ees}{\end{subequations}}
\def\bal#1\eal{\begin{align}#1\end{align}}
\newcommand{\ov}{\overline}

\newcommand{\nn}{\nonumber\\}
\newcommand{\bfi}{\begin{figure}}
\newcommand{\efi}{\end{figure}}
\newcommand{\bc}{\begin{center}}
\newcommand{\ec}{\end{center}}

\newcommand{\p}{\partial}

\newcommand{\vphi}{\varphi}
\newcommand{\vphic}{\ov{\varphi}}
\newcommand{\vphia}{|\varphi|}

\begin{document}
\title{Vortices in Maxwell-Higgs models with a global factor}
\author{I. Andrade}\email{andradesigor0@gmail.com}\affiliation{Departamento de F\'\i sica, Universidade Federal da Para\'\i ba, 58051-970 Jo\~ao Pessoa, PB, Brazil}
\author{M.A. Marques}\email{marques@cbiotec.ufpb.br}\affiliation{Departamento de Biotecnologia, Universidade Federal da Para\'\i ba, 58051-900 Jo\~ao Pessoa, PB, Brazil}\affiliation{Departamento de F\'\i sica, Universidade Federal da Para\'\i ba, 58051-970 Jo\~ao Pessoa, PB, Brazil}
\author{R. Menezes}\email{rmenezes@dcx.ufpb.br}\affiliation{Departamento de Ci\^encias Exatas, Universidade Federal
da Para\'{\i}ba, 58297-000 Rio Tinto, PB, Brazil}
\affiliation{Departamento de F\'\i sica, Universidade Federal da Para\'\i ba, 58051-970 Jo\~ao Pessoa, PB, Brazil}

\date{\today}

\begin{abstract}
This paper deals with planar vortices in a generalized model that presents a global factor which depends on the scalar field in the Nielsen-Olesen Lagrange density. We show that the system supports a first order framework. Contrary to what occurs with kinks in the line, planar vortices require the presence of constraints that brings modifications into the first order equations. Novel features are unveiled, such as finite energy configurations with infinite energy density at the origin and the presence of parameters that modifies the energy, keeping the solutions unchanged.
\end{abstract} 

\maketitle

In high energy physics, topological strucutres appear under the action of scalar field models \cite{manton,vachaspati}. The most common objects with these nature are kinks, vortices and monopoles. Kinks are the simplest ones and arise in the presence of a single real scalar field in the line. Vortices appear in the plane, with a complex scalar field coupled to a gauge field under an $U(1)$ local symmetry \cite{NO}. To study monopoles, one must consider a triplet of scalar fields coupled to a gauge field under a $SU(2)$ symmetry in the tridimensional space \cite{thooft,polyakov}.

The usual path to study the aforementioned structures is by taking Lagrange densities with the sum of the dynamical terms associated to each field and the potential that dictates the self-interaction of the scalar field. There are other possibilities, the so-called noncanonical models, to investigate topological objects. The inspiration comes from cosmology, in the context of inflation \cite{kinf,cosm1,cosm2}. There, generalized models can be used to drive the inflationary evolution without the presence of a potential and, also, as a tentative solution to the problem of the cosmic coincidence.

In the context of noncanonical models, we have previously investigated how the inclusion of a global factor depending on the scalar field modifies the standard kink scenario in Ref.~\cite{fL}. Interestingly, the model supports two situations, related to the function chosen. In the first case, the solution is the same of the standard model, but the energy density is different. In the second case, one may choose a global function that leads to the same energy density, but with nonstandard solutions. In Ref.~\cite{p1}, the authors have studied extensions of this idea in which the global factor is a function of the potential.

In this letter, we investigate how the inclusion of a global factor in the Lagrange density modifies the Nielsen-Olesen vortex \cite{NO}, using a first order framework compatible with the equations of motion \cite{bogopaper,godvortex}. Before dealing with planar vortices, we take a brief look at the kink scenario in $(1,1)$ spacetime dimension with metric tensor $\eta_{\mu\nu}=\text{diag}(+,-)$, considered in Ref.~\cite{fL}. There, the action is
\be\label{skink}
S = \int dx\,dt\,f(\phi)\left(\frac12\partial_\mu\phi\partial^\mu\phi - V(\phi)\right),
\ee
in which $\phi$ is a real scalar field. We work with natural units $(\hbar=c=1)$, so the dimensions of the involved quantities are given in terms of a power-law form of the energy, such that the powers are dim$(x_\mu)=-1$ and dim$(V(\phi))=2$; both field and the function $f(\phi)$ are dimensionless. The equation of motion for static configurations is 
\be\label{eomkink}
f(\phi)\left(\phi^{\prime\prime} - V_\phi\right) = -f_\phi\left(\frac12{\phi^{\prime}}^2-V(\phi)\right),
\ee
where the prime denotes the derivative with respect to spatial coordinate $x$, $V_\phi=dV/d\phi$ and $f_\phi=df/d\phi$. Coincidently, the right hand side of this equation vanishes under the stressless condition, which is associated to the stability of the solutions against contractions and dilations \cite{trilogia1}. So, one obtains $\phi^{\prime\prime} = V_\phi$, which is the very same equation of the standard case, in which $f(\phi)=1$. It is also possible to integrate this equation into first order, in the form ${\phi^{\prime}}^2=2V(\phi)$. The energy density associated to the solutions of this equation is
\be\label{rhokink}
\rho(x) = 2f(\phi(x)) V(\phi(x)).
\ee
Thus, even though the function $f(\phi)$ does not modify the solution, it appears as a global factor in the energy density.

In this work, we deal with vortex structures following a similar idea of Ref.~\cite{fL}. Here, however, we must consider models in $(2,1)$ spacetime dimensions with a complex scalar field coupled to a gauge one through an $U(1)$ local symmetry. Considering the metric tensor to be $\eta_{\mu\nu}=\text{diag}(+,-,-)$, we take the action
\be\label{model}
S = \int dt\,d^2x\, f(\vphia)\left(-\frac{1}{4}F_{\mu\nu}F^{\mu\nu} +\ov{D_\mu\vphi}D^\mu\vphi - U(\vphia)\right).
\ee
Here, $F_{\mu\nu}=\p_\mu A_\nu - \p_\nu A_\mu$ denotes the electromagnetic tensor and $D_\mu=\p_\mu +ieA_\mu$ represents the derivative with the minimal coupling. The dimensions of the quantities in this case are dim$(\vphi)=$ dim$(A_\mu)=$ dim$(e)=1/2$ and dim$(U(\vphia))=3$. The function $f(\vphia)$ is non-negative and dimensionless. The equations of motion for the above Lagrange density are
\bes\label{eom}
\bal\label{eomphi}
&D_\mu\left(f(\vphia)D^\mu\vphi\right) = \frac{\vphi}{2\vphia}\bigg(-\frac{f_{\vphia}}{4}F_{\mu\nu}F^{\mu\nu}\nn
	&+f_{\vphia}\ov{D_\mu\vphi}D^\mu\vphi -f(\vphia)U_{\vphia} -U(\vphia)f_{\vphia}\bigg),\\
\label{eomA}&\p_\mu\left(f(\vphia)F^{\mu\nu}\right) = J^\nu,
\eal
\ees
where $J_\mu=ief(\vphia)(\vphic D_\mu\vphi -\vphi\ov{D_\mu\vphi})$ is a $3$-current. The energy-momentum tensor associated to the Lagrange density Eq.~\eqref{model} is given by
\be\label{tmunu}
\begin{aligned}
T_{\mu\nu} &= f(\vphia)\bigg(F_{\mu\lambda}\tensor{F}{^\lambda_\nu} +\frac14\eta_{\mu\nu}F_{\lambda\sigma}F^{\lambda\sigma} +\eta_{\mu\nu}U(\vphia)\\
	&+\ov{D_\mu\vphi}D_\nu\vphi +D_\mu\vphi\ov{D_\nu\vphi} -\eta_{\mu\nu}\ov{D_\lambda\vphi}D^\lambda\vphi\bigg).
\end{aligned}
\ee
By considering static configurations and setting $\nu=0$ in Eq.~\eqref{eomA}, one can see it is satisfied by $A_0=0$. In this situation, there is no electric field associated to the structure. To investigate vortex solutions, we take the usual ansatz
\be\label{ansatz}
\vphi = g(r)e^{in\theta} \quad\text{and}\quad \vec{A} = \frac{\hat{\theta}}{er}\left(n -a(r)\right),
\ee
where $n=\pm1,\pm2,\ldots$ is the vorticity, with the boundary conditions $g(0) = 0$, $g(\infty) \to v$, $a(0) = n$ and $a(\infty) \to 0$.

Even though the vortex is electrically neutral, with the absence of the electric field, it engenders a magnetic field which is given by $B=-F^{12}$. With the above ansatz, it becomes
\be\label{mfield}
B = -\frac{a^\prime}{er},
\ee
where the prime now denotes the derivative with respect to $r$. By integrating this and considering the aforementioned boundary conditions, we obtain a quantized magnetic flux, in the form $\Phi = 2\pi n/e$. Notice that this result is general and does not depend explicitly on the functions $f(\vphia)$ and $U(\vphia)$.

The equations of motion \eqref{eom} with the ansatz become
\bes\label{eomansatz}
\begin{align}
&\frac{1}{r}\left(rfg^\prime\right)^\prime -\frac{a^2gf}{r^2} -\frac12f_{\vphia}\left({g^\prime}^2 +\frac{a^2g^2}{r^2}\right)\nonumber\\
&-\frac{f_{\vphia}{a^\prime}^2}{4e^2r^2} -\frac12fU_{\vphia} -\frac12Uf_{\vphia}= 0,\\
&\left(\frac{fa^\prime}{er}\right)^\prime -\frac{2eag^2f}{r} = 0,
\end{align}
\ees
The energy density can be calculated with the $T_{00}$ component in Eq.~\eqref{tmunu}; it has the form
\be\label{rho}
\rho = f(g)\left(\frac{{a^\prime}^2}{2e^2r^2} +{g^\prime}^2 +\frac{a^2g^2}{r^2} +U(g)\right).
\ee
The above expression shows that, as in the Lagrange density \eqref{model}, the function $f(g)$ appears as a global factor in the energy density. The equations \eqref{eomansatz} are of second order and present couplings between the functions $a(r)$ and $g(r)$. This makes them be hard to solve. In order to find first order equations, we follow the lines of Ref.~\cite{godvortex}. There, the authors show that first order equations arise with the stressless condition, $T_{ij}=0$; see also Refs.~\cite{vega,atmaja1,atmaja2}. They have the form
\bes\label{fo}
\bal\label{fog}
g^\prime &= \pm\frac{ag}{r},\\
\label{foB}-\frac{a^\prime}{er} &= \pm\sqrt{2U(g)}.
\eal
\ees
The solutions that satisfy these equations are stable under contractions and dilations. Notice the above equations has two signs. The upper/lower sign represents positive/negative vorticity, related by the change $a\to-a$. So, the configurations with negative vorticity can be easily found by knowing the positive ones.

We remark that the function $f(\vphia)$ does not appear explicitly in the above first order equations. This occurs because it is a global factor in the Lagrange density \eqref{model}. At first glance, one may think that, since these equations are identical to the standard case, $(f(\vphia)=1)$, the solutions are the very same of the standard model as in the kink scenario in Eq.~\eqref{skink}. Nevertheless, this is not true because the vortex configurations are governed by the equations of motion \eqref{eomansatz} and one must ensure that the above equations are compatible with them. This associates the function $U(\vphia)$ to $f(\vphia)$ through the equation
\be\label{vinc}
\frac{d}{d\vphia}\left(\sqrt{2f^2(\vphia)U(\vphia)}\right) = -2e\vphia f(\vphia).
\ee
By solving the above equation for $U(\vphia)$, we get
\be\label{pot}
U(\vphia) = \frac{2e^2}{f^2(\vphia)}\left(\int d\vphia\,\vphia f(\vphia)\right)^2.
\ee
So, for a given $f(\vphia)$ there is a corresponding $U(\vphia)$ that defines the starting model in Eq.~\eqref{model}. Notice the presence of an integral in this equation; it gives rise to a constant that may be related to the symmetry breaking of the model.

The Nielsen-Olesen model, investigated in Ref.~\cite{NO}, is recovered for $f(\vphia)=1$. Indeed, the potential \eqref{pot} simplifies to
\be\label{ustd}
U_{std}(\vphia) = \frac{e^2}{2}\left(v^2-\vphia^2\right).
\ee
The first order equation \eqref{fog} does not change, and Eq.~\eqref{foB} becomes
\be\label{foastd}
-\frac{a^\prime}{er} = \pm e\left(v^2-g^2\right).
\ee
In the latter two equations, $v$ is a symmetry-breaking parameter that appears as an integration constant in Eq.~\eqref{pot}.

For a general $f(\vphia)$, an interesting feature is that, regardless the form of $U(\vphia)$, Eq.~\eqref{fog} allows one to show that, near the origin, the function that controls the scalar field behaves as $g(r\approx 0)\propto r^{|n|}$.  We also highlight that the above first order equations leads to solutions that minimizes the energy of the system \cite{bogopaper}.

One can introduce an auxiliary function, $W(a,g)$, such that
\be\label{W}
\begin{split}
W(a,g) &= -\frac{af(g)}{e}\sqrt{2U(g)},\\
	   &= 2a\left(\int d\vphia\,\vphia f(\vphia)\right),	
\end{split}
\ee
in which Eq.~\eqref{pot} was used to obtain the expression in the latter line. The above function allows us to write the energy density in terms of a total derivative, as $\rho=\pm W^\prime/r$. By integrating it, one gets the minimum energy in the following form
\be\label{energy}
E = 2\pi\left|W(0,v) -W(n,0)\right|.
\ee
Notice that this energy depends only on the boundary conditions associated to the solutions. Thus, if the potential is written as \eqref{pot}, one can calculate the energy without knowing the explicit form of the solutions. Note that the change $f(g)\to\alpha f(g)$, with $\alpha$ being a dimensionless parameter, modifies the energy due to the form of Eq.~\eqref{W}, despite the invariance of the potential \eqref{pot} and of the first order equations \eqref{fo} through this transformation.

The first order equations \eqref{fo} can be used to rewrite the energy density \eqref{rho} exclusively in terms of the solutions $a(r)$ and $g(r)$, as
\be\label{rhosol}
\rho(r) = 2f(g(r))\left(\frac{a^2(r)g^2(r)}{r^2} +U(g(r))\right).
\ee
Note that the function $f(g)$ appears as a global factor, similarly as in the kink scenario in Eq.~\eqref{rhokink}, but as we have commented before, the solutions are not the standard ones.

\begin{figure}[t!]
\centering
\includegraphics[width=4.2cm,trim={0.6cm 0.7cm 0 0},clip]{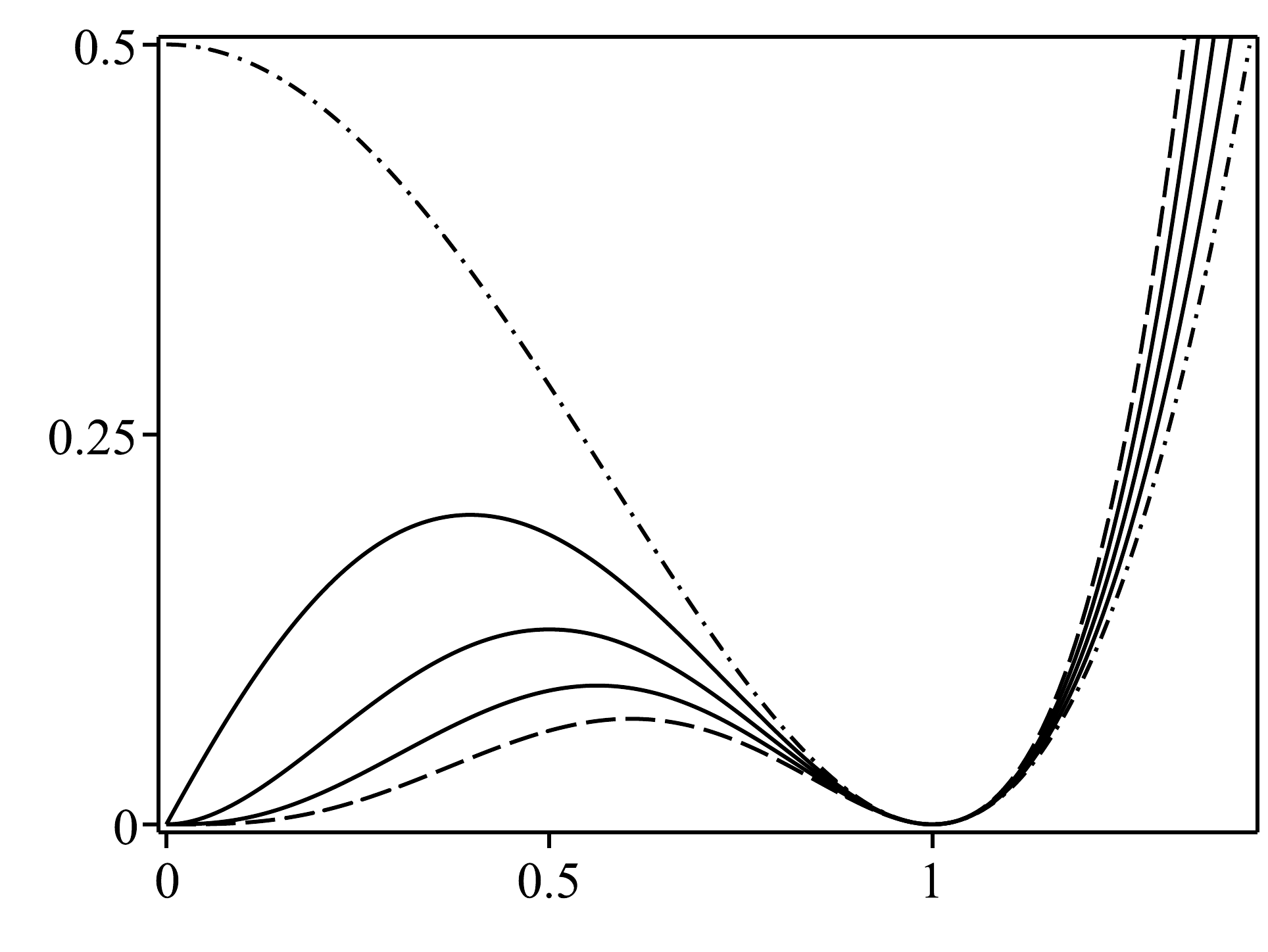}
\includegraphics[width=4.2cm,trim={0.6cm 0.7cm 0 0},clip]{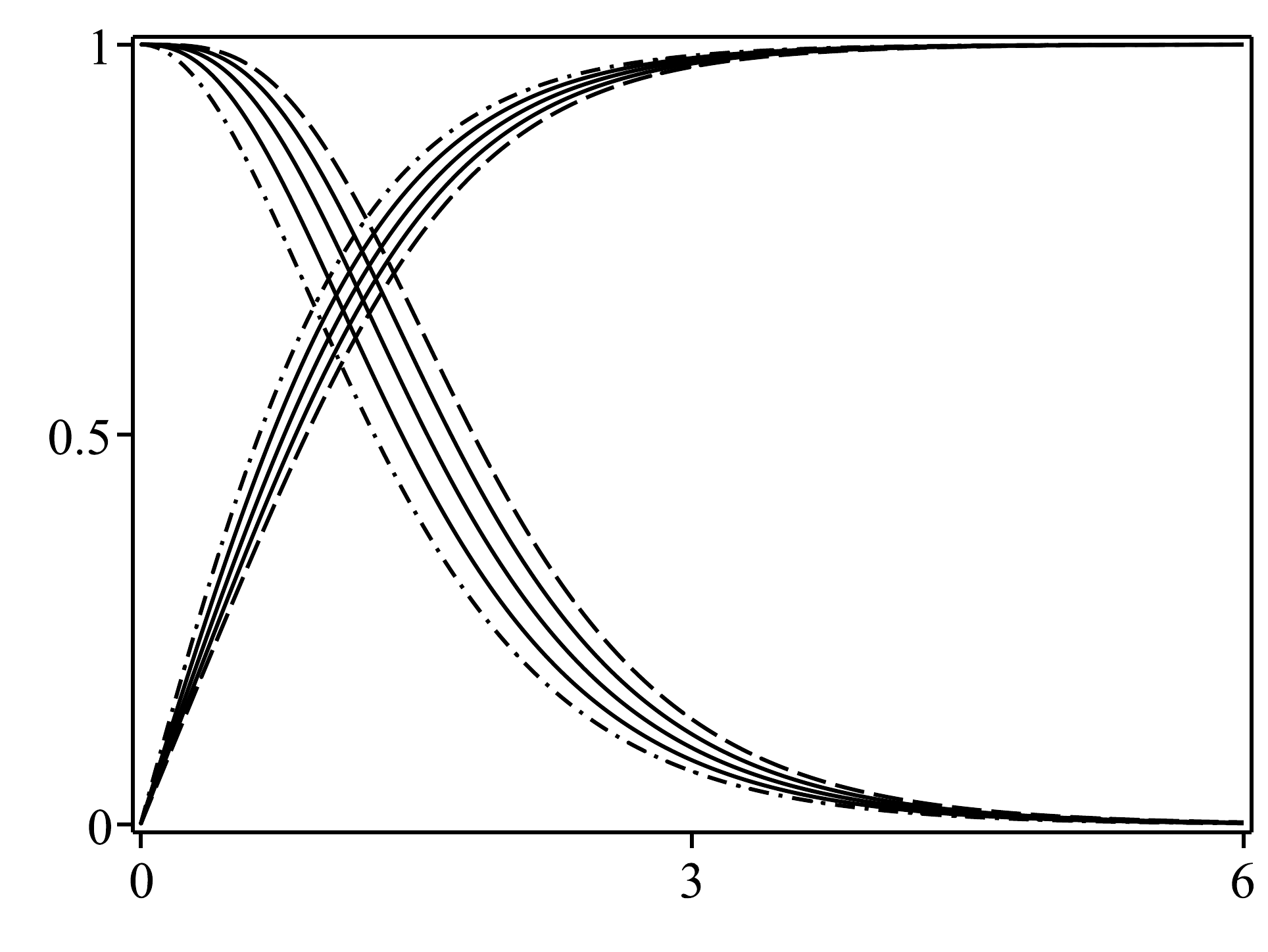}
\caption{The potential \eqref{pot1} (left) and the associated solutions $a(r)$ (descending lines) and $g(r)$ (ascending lines) (right) for $e,v,n=1$ and some values of $l$. The dash-dotted lines denote $l=0$, the solid ones stand for $l=0.5,1,1.5$ and the dashed ones represent $l=1.999$.}
\label{fig1}
\end{figure}

Next, we consider novel configurations that arise with the function $f(\vphia)$ in the above first order framework. The first possibility that we present is the one with a power-law function, in the form
\be\label{f1}
f(\vphia) = c\left(\frac{\vphia}{v}\right)^{-l},
\ee
where $c$ is a non-negative dimensionless parameter and $l\in[0,2)$ to obtain finite energy configurations. For positive $l$, it diverges when the field is null. The potential is given by Eq.~\eqref{pot}, which leads us to
\be\label{pot1}
U(\vphia) = \frac{2e^2v^4}{\left(2-l\right)^2}\left(\frac{\vphia}{v}\right)^{2l}\left(1 -\left(\frac{\vphia}{v}\right)^{2-l}\right)^2,
\ee
where $v$ is a parameter involved in the symmetry breaking of the model that arises as an integration constant. For $l=0$ we recover the potential of the standard model in Eq.~\eqref{ustd}, such that the point $\vphia=0$ determines a local maximum and $\vphia=v$ leads to a set of minima. For $l>0$, the point $\vphia=v$ is a set of minima and $\vphia=0$ becomes a minimum point instead of a maximum one. The above potential can be seen in Fig.~\ref{fig1} for $e,v=1$ and some values of $l$.

\begin{figure}[t!]
\centering
\includegraphics[width=4.2cm,trim={0.6cm 0.7cm 0 0},clip]{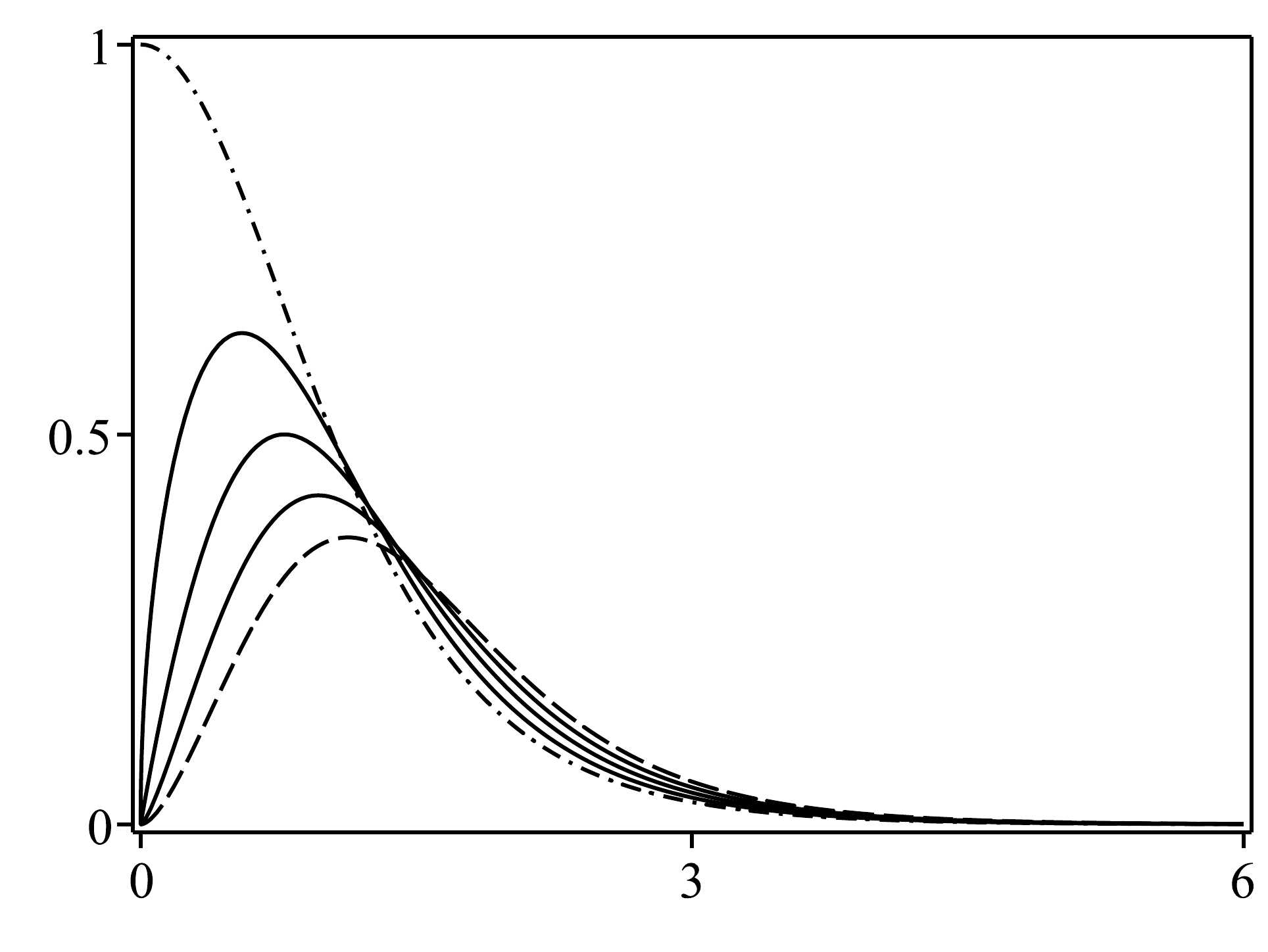}
\includegraphics[width=4.2cm,trim={0.6cm 0.7cm 0 0},clip]{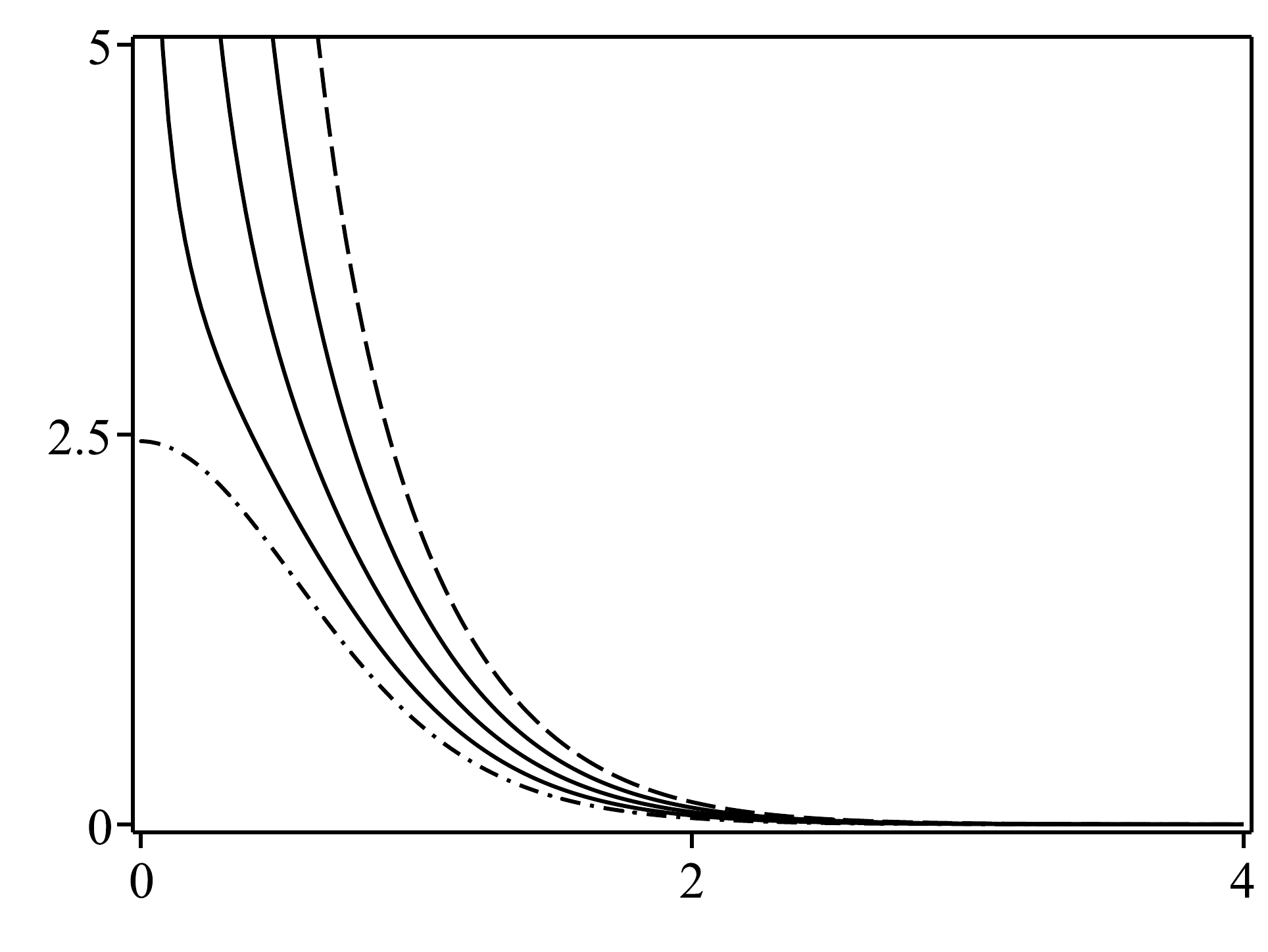}
\caption{The magnetic field \eqref{mfield} (left) and the energy density \eqref{rhosol} (right) associated to the potential in Eq.~\eqref{pot1} for $e,v,n,c=1$ and some values of $l$. The dash-dotted lines denote $l=0$, the solid ones stand for $l=0.5,1,1.5$ and the dashed ones represent $l=1.999$.}
\label{fig2}
\end{figure}

To calculate the solutions, we must solve the system of differential equations formed by Eq.~\eqref{fog}, which remains unchanged, and Eq.~\eqref{foB}, which leads us to
\be
-\frac{a^\prime}{er} = \pm\frac{2ev^2}{2-l}\left(\frac{g}{v}\right)^l\left(1 -\left(\frac{g}{v}\right)^{2-l}\right).
\ee
The right hand side of the above equation also describes the magnetic field, since it is given by \eqref{mfield}. We were not able to find the analytical solutions for this system. However, near the origin, one can show that $n-a(r\approx0)\propto r^{2+|n|l}$. On the other hand, the solutions behave asymptotically as 
\bes\label{asymptstd}\bal
v-g(r>>0) &\propto \frac{1}{\sqrt{r}}\exp\left(-\sqrt{2}evr\right),\\
a(r>>0)&\propto \sqrt{r}\exp\left(-\sqrt{2}evr\right),
\eal
\ees
exactly as in the standard case, $f(\vphia)=1$. In Fig.~\ref{fig1}, we plot the solutions for $e,v,n=1$ and some values of $l$. The auxiliary function \eqref{W} is
\be
W(a,g) = -\frac{2cv^2a}{2-l}\left(1 -\left(\frac{g}{v}\right)^{2-l}\right).
\ee
Thus, the energy in Eq.~\eqref{energy} is given by $E=4\pi cv^2|n|/(2-l)$. The energy density is given by Eq.~\eqref{rhosol} with the function \eqref{f1} and potential \eqref{pot1}. One can show that the behavior of the energy density at the origin depends on the vorticity. Indeed, for $|n|=1$, it behaves as $\rho(r\approx0)\propto r^{-l}$; so, at the origin, it is finite and non-null for $l=0$ and infinite otherwise. In the case $|n|\neq 1$, it depends of the value of $l$: for $l<1-1/|n|$ it behaves as $\rho(r\approx0)\propto r^{|n|l}$ and for $l\geq 1-1/|n|$ it behaves as $\rho(r\approx0)\propto r^{|n|(2-l)-2}$. This leads us to three situations. The first one occurs for $l=0$ and $l=2-2/|n|$, in which the energy density is finite and non-null at $r=0$. The second one appears for $0<l<2-2/|n|$, which leads to null energy density at $r=0$. The third one leads to a divergence at $r=0$ and arises for $l>2-2/|n|$. Although a divergence at the origin may appear in the energy density, the energy is finite. One can see the magnetic field \eqref{mfield} and energy density \eqref{rhosol} associated to the potential in Eq.~\eqref{pot1} for $e,v,n,c=1$ and some values of $l$ in Fig.~\ref{fig2}.

The previous model supports vortex solutions with finite energy and infinite energy density at the origin. To circumvent this divergence, we introduce a novel model, described by
\be\label{f2}
f(\vphia) = c\left(1-l\sin\left(\frac{\pi \vphia^2}{v^2}\right)\right).
\ee
In this model, $l$ is a real parameter defined in the interval $l\in(-1,1)$ to obtain positive values of $f(\vphia)$. In this case, the potential \eqref{pot} becomes
\be\label{pot2}
\begin{aligned}
U(\vphia) &= \frac{e^2v^4}{2}\left(1-l\sin\left(\frac{\pi \vphia^2}{v^2}\right)\right)^{-2}\\
	&\times\left(1 -\frac{\vphia^2}{v^2} -\frac{l}{\pi}\left(1 +\cos\left(\frac{\pi \vphia^2}{v^2}\right)\right)\right)^2,
\end{aligned}
\ee
in which the integration constant was chosen to allow for the presence of  a set of minima at $\vphia=v$. This potential engender a point that is always present, regardless the value of $l$, which we call $\vphia=\tilde{g}$. It is given by the algebraic equation
\be
1+\cos\left(\frac{\pi \tilde{g}^2}{v^2}\right) = \pi\left(1 -\frac{\tilde{g}^2}{v^2}\right)\sin\left(\frac{\pi \tilde{g}^2}{v^2}\right).
\ee
Unfortunately, we were not able to solve it analytically. By using numerical procedures, we have found $\tilde{g}=0.5079$, such that $U(\tilde{g}) =0.2753$.
\begin{figure}[t!]
\centering
\includegraphics[width=4.2cm,trim={0.6cm 0.7cm 0 0},clip]{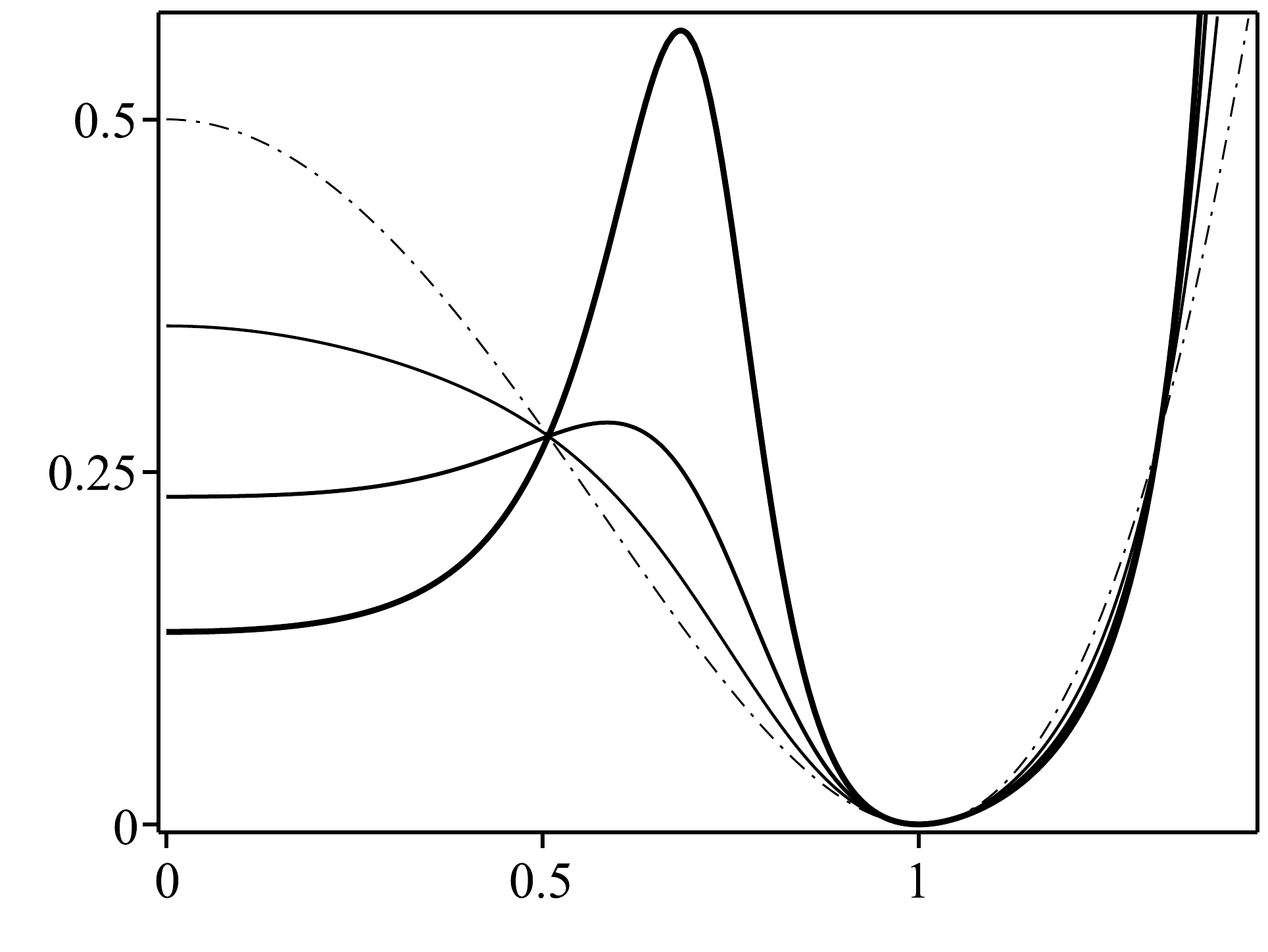}
\includegraphics[width=4.2cm,trim={0.6cm 0.7cm 0 0},clip]{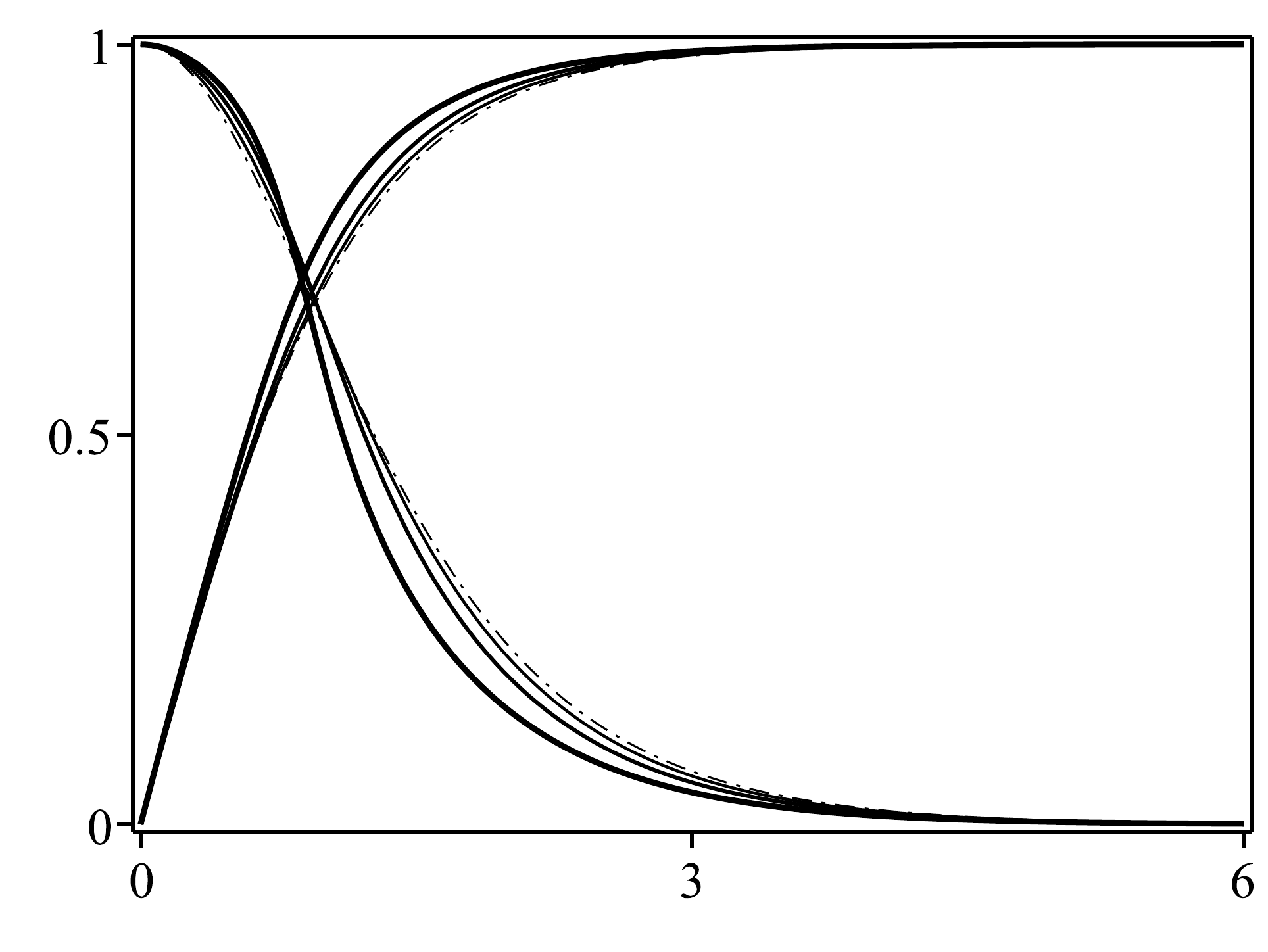}
\caption{The potential \eqref{pot2} (left) and its associated solutions $a(r)$ (descending lines) and $g(r)$ (ascending lines) (right), for $e,v,n=1$ and some values of $l$. The dash-dotted lines stand for $l=0$ and the solid ones represent $l=0.25,0.5,0.75$, with the thickness increasing with $l$.}
\label{fig3}
\end{figure}
In this case, the first order equations that we must solve are Eq.~\eqref{fog} and \eqref{foB}. The latter one, for the current model, takes the form
\be
\begin{aligned}
-\frac{a^\prime}{er} &= \pm ev^2\left(1-l\sin\left(\frac{\pi g^2}{v^2}\right)\right)^{-1}\\
	&\times\left(1 -\frac{g^2}{v^2} -\frac{l}{\pi}\left(1 +\cos\left(\frac{\pi g^2}{v^2}\right)\right)\right).
\end{aligned}
\ee
Even though the function \eqref{f2} introduces modifications in the potential, the first order equations lead us to the very same behavior of the standard case, $f(\vphia)=1$, both at the origin, $n-a(r\approx0)\propto r^2$, and at infinity, which can be found in Eq.~\eqref{asymptstd}. In Fig.~\ref{fig3}, one can see the potential \eqref{pot2} and the solutions of the aforementioned first order equations for $e,v,n=1$ and some non-negative values of $l$. The configurations that arise with negative $l$ are very similar to the ones in the standard case, described by Eqs.~\eqref{ustd} and \eqref{foastd}, so we do not deal with them.

\begin{figure}[t!]
\centering
\includegraphics[width=4.2cm,trim={0.6cm 0.7cm 0 0},clip]{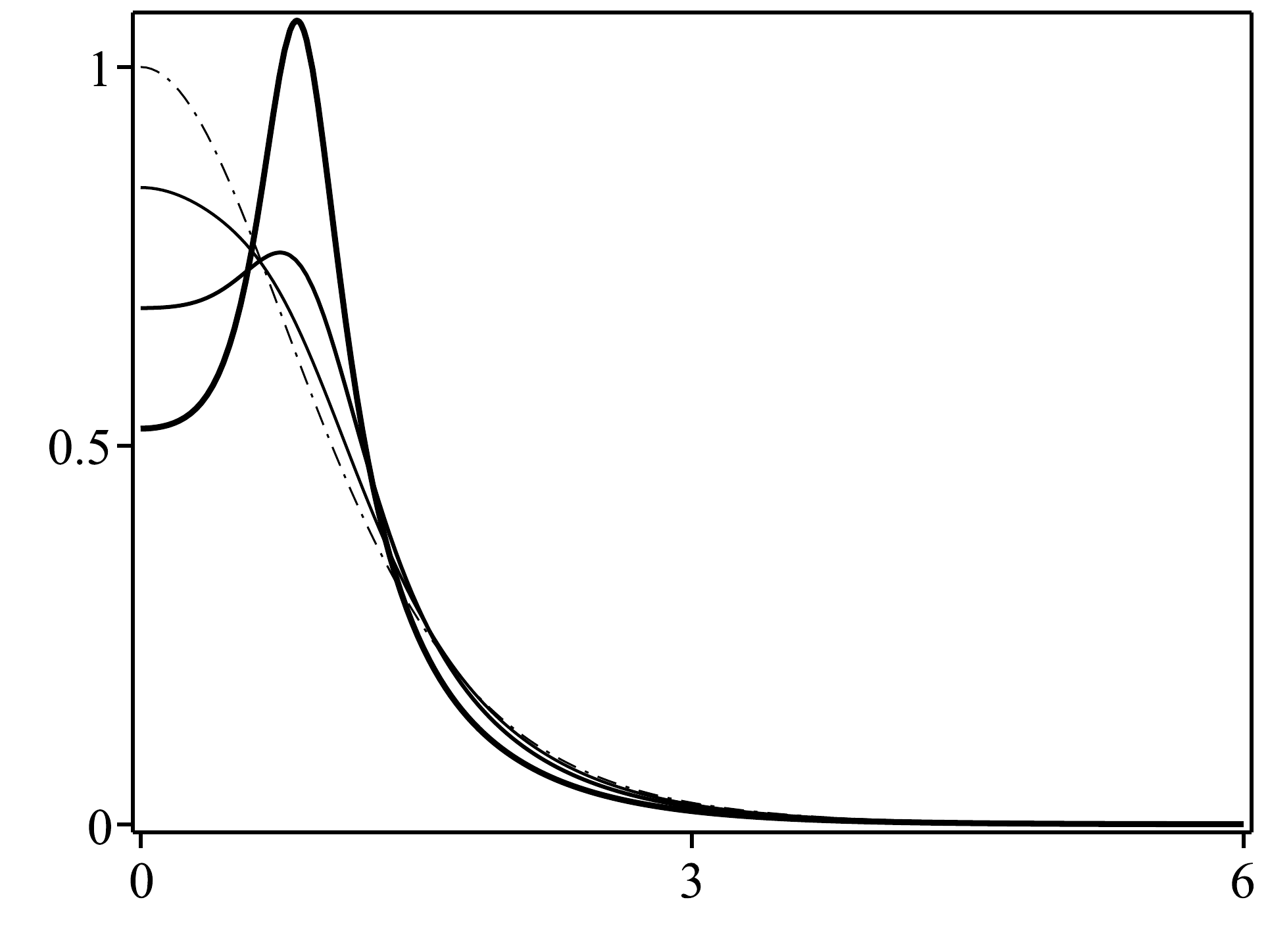}
\includegraphics[width=4.2cm,trim={0.6cm 0.7cm 0 0},clip]{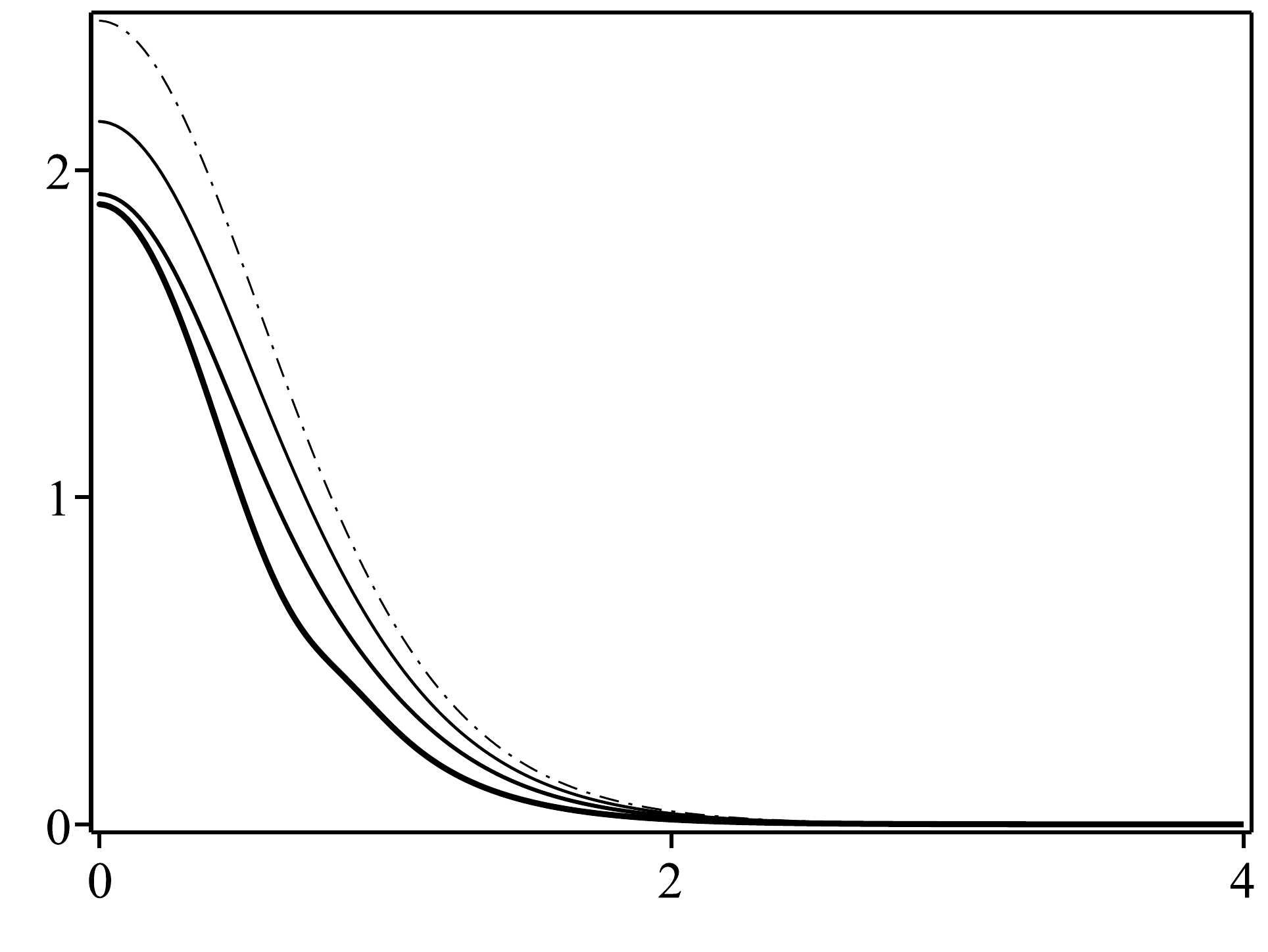}
\caption{The magnetic field \eqref{mfield} (left) and the energy density \eqref{rhosol} (right) associated to the potential \eqref{pot2}. The dash-dotted lines stand for $l=0$ and the solid ones represent $l=0.25,0.5,0.75$, with the thickness increasing with $l$.}
\label{fig4}
\end{figure}

The auxiliary function \eqref{W} is
\be
W(a,g) = -cv^2a\left(1 -\frac{g^2}{v^2} -\frac{l}{\pi}\left(1 +\cos\left(\frac{\pi g^2}{v^2}\right)\right)\right),
\ee
such that the energy \eqref{energy} is $E=2cv^2|n|(\pi-2l)$. The energy density is given by  \eqref{rhosol} with the function \eqref{f2} and potential \eqref{pot2}. In Fig.~\ref{fig4}, one can see the magnetic field \eqref{mfield} and the energy density for $e,v,n,c=1$ and some values of $l$. In this case, the energy density is regular, without divergency as in the previous model.

An interesting situation arises in the case in which the function $f(\vphia)$ depends on the potential. Here, we take the following power-law form
\be\label{f3}
f(\vphia)=\lambda U^l(\vphia),
\ee
such that the dimension of the parameter $\lambda$ is dim$(\lambda)=-3l$ and $l$ is a real parameter, in principle. In this situation, the constraint in Eq.~\eqref{vinc} simplifies to
\be
\frac{d}{d\vphia}\left(\sqrt{2U^{2l+1}(\vphia)}\right) = -2e\vphia U^l(\vphia).
\ee
From this equation, we see that the case $l=-1/2$ leads to $U(\vphia)=0$. We then consider $l>-1/2$ and solve the above equation, which leads to
\be\label{pot3}
U(\vphia) = \frac{e^2\left(v^2-\vphia^2\right)^2}{2\left(2l+1\right)^2},
\ee
in which the symmetry breaking parameter $v$ arises as an integration constant. Notice that the case $l=0$ leads to the potential of the standard case, with the well-known form in Eq.~\eqref{ustd}, and $f(\vphia)=\lambda$.

To calculate the solutions, we must solve Eq.~\eqref{fog}, which remains unchanged, and Eq.~\eqref{foB}, whose form is now given by
\be
-\frac{a^\prime}{er} = \pm\frac{e\left(v^2-g^2\right)}{2l+1}.
\ee
This problem is very similar to the standard case $(l=0)$. In fact, one can show that this solutions are related with the standard ones by the change $e\to \sqrt{1+2l}\,e$. Near the origin, it behaves exactly as in the standard case. However, the asymptotic behavior is modified by the parameter $l$. One can do the aforementioned change in Eq.~\eqref{asymptstd} to obtain
\bes
\bal
v-g(r>>0) &\propto \frac{1}{\sqrt{r}}\exp\left(-\sqrt{\frac{2}{2l+1}}\,evr\right),\\
a(r>>0)&\propto \sqrt{r}\exp\left(-\sqrt{\frac{2}{2l+1}}\,evr\right).
\eal
\ees
This makes the energy density behave asymptotically as
\be
\rho(r>>0) \propto \frac{1}{r^{l+1}}\exp\left(-\frac{2\sqrt{2}ev(l+1)r}{\sqrt{2l+1}}\right).
\ee
Note that, even though the change $e\to \sqrt{1+2l}\,e$ relates the solutions with the ones of the standard case, the same does not occur with the energy density.
 
We do not display the potential in Eq.~\eqref{pot3} and its associated solutions because they are similar to the dash-dotted lines found in Fig.~\ref{fig1}. We can take advantage of the auxiliary function in Eq.~\eqref{W} to show that
\be
W(a,g) = -\frac{\lambda e^{2l}a\left(v^2-g^2\right)^{2l+1}}{2^l\left(2l+1\right)^{2l+1}}.
\ee
The above expression leads us to an energy, according to Eq.~\eqref{energy}, in the form $E=2^{1-l}\pi\lambda e^{2l}v^{4l+2}|n|/(2l+1)^{2l+1}$. The profile of both the magnetic field \eqref{mfield} and the energy density \eqref{rhosol} for this model is similar to the dash-dotted lines found in Fig.~\ref{fig2}, so we do not plot them here.

In this paper, we investigated how a global factor depending on the scalar field modifies the vortex configurations in the Nielsen-Olesen Lagrange density \cite{NO}. We get inspiration from Ref.~\cite{fL}, in which we studied a similar approach in the kink scenario, in $(1,1)$ dimensions, and use a first order formalism to describe the vortex solutions with minimum energy. Whilst the presence of the aforementioned global function did not modified the differential equations that give rise to the kink profile, it changes the ones involved in the vortex scenario. This occurs because the model in Eq.~\eqref{model} requires constraints to ensure the compatibility of the first order equations with the equations of motion.

To illustrate our general model, we have first considered a power-law function of the field, and showed that it leads to finite energy vortex configurations, even though the energy density may be infinity at the origin for some specific range of the parameters. In the second model, we have taken $f(\vphia)$ as a trigonometric function. It presents a parameter $l$ that controls the behavior of the magnetic field near the origin, making hole at this point as $l$ increases. Even though the magnetic field is modified, the flux remain constant, $\Phi=2\pi n/e$. The third model investigated was a case in which the function $f(\vphia)$ is a power-law form of the potential. In this situation, the solutions may related to the ones of the standard case by a rescale in the charge $e$. Notwithstanding that, the energy density does not become the same of the standard case with this change.

As perspectives, one may investigate the case of additional symmetries, following the lines of Refs.~\cite{schaposnik1,schaposnik2}, which include an extra $U(1)$ local symmetry to accommodate the so-called hidden sector. One may also study the presence of the aforementioned global factor in models whose enhanced symmetry is used to give rise to multilayered structures \cite{prr}. These possibilities takes into account configurations with topological nature. One can also investigate how the global factor in the Lagrange density modifies nontopological objects, such as Q-balls, and their properties: energy, charge and stability \cite{coleman,tdlee,cervero,kumar,qball}. Some of these issues are under consideration and we hope to report on them elsewhere.

\acknowledgements{We would like to thank Dionisio Bazeia for the discussions that have contributed to this paper. The work is supported by the Brazilian agencies Conselho Nacional de Desenvolvimento Cient\'ifico e Tecnol\'ogico (CNPq), grants Nos. 140490/2018-3 (IA) and 306504/2018-9 (RM), and by Paraiba State Research Foundation (FAPESQ-PB) grants Nos. 0003/2019 (RM) and 0015/2019 (MAM).}

\end{document}